\documentclass[showpacs,showkeys,amsmath,amssymb,10pt,superscriptaddress,reprint, preprintnumbers,nofootbib,notitlepage, groupdaddress,showabstract,,aps,prl,twocolumn]{revtex4-1}

\pdfoutput=1 
\usepackage[utf8]{inputenc}
\usepackage[english]{babel}
\usepackage{bm}
\usepackage{graphicx}
\usepackage{dcolumn}
\usepackage{pbox}
\usepackage{epsfig}
\usepackage[dvipsnames]{xcolor}
\usepackage{slashed}
\usepackage{amssymb}
\usepackage{ mathrsfs }
\usepackage{color}
\usepackage[font=small]{caption}
\usepackage[font=small]{subcaption}
\usepackage{url}
\definecolor{DarkBlue}{rgb}{0.7, 0.4, 1} 
\definecolor{Blue}{rgb}{0, 0.8, 0} 
\definecolor{MyLightBlue}{rgb}{0.5,0.7,1.9}
\definecolor{MyGreen}{rgb}{0.0,0.2, 0.0}
\definecolor{MyBrickRed}{rgb}{0, 0.5, 0.2}
\RequirePackage{hyperref}
\hypersetup{colorlinks, citecolor=Blue,linkcolor=DarkBlue, urlcolor=Green}

\newcommand{\bea}{\begin{eqnarray}}
\newcommand{\eea}{\end{eqnarray}}

\makeatletter

\renewcommand\@makecaption[2]{%
  \par
  \vskip\abovecaptionskip
  \begingroup
  
   \small\rmfamily
    \begingroup
     \samepage
     \flushing
     \let\footnote\@footnotemark@gobble
     \@make@capt@title{#1}{#2}\par
    \endgroup
  \endgroup
  \vskip\belowcaptionskip
}
\makeatother

\DeclareUnicodeCharacter{2212}{-}
\begin{document}
\title{Enhancement of the Higgs decay into a $Z^\prime$ pair in models with $U(1)_X$ gauge symmetry }

\preprint{OU-HET-1228}

\author{ShivaSankar K.A.}
\email{a-shiva@particle.sci.hokudai.ac.jp}
\affiliation{Department of Physics, Hokkaido University, Sapporo 060-0810, Japan}
\author{Arindam Das}
\affiliation{Institute for the Advancement of Higher Education, Hokkaido University, Sapporo 060-0817, Japan}
\affiliation{Department of Physics, Hokkaido University, Sapporo 060-0810, Japan}
\author{Kei Yagyu}
\affiliation{Department of Physics, Osaka University, Toyonaka, Osaka 560-0043, Japan}

\begin{abstract}
We discuss the Higgs phenomenology in models with a new $U(1)_X$ gauge symmetry including the $U(1)_{B-L}$ scenario, where three right-handed neutrinos are inevitably introduced due to the gauge anomaly cancellations. We find that the decay branching ratio of the discovered Higgs boson into a pair of new massive gauge bosons ($Z'$) can significantly be enhanced in the Dirac neutrino case as compared with the Majorana case for a fixed value of the new gauge coupling and the mass of $Z'$ under constraints from current experimental data.  
Because of such an enhancement, the Dirac case can indirectly be discriminated from the Majorana case via the Higgs decay.
\end{abstract}
\maketitle
\noindent
{\textbf{Introduction}.--}
The origin of neutrino masses has long been a mystery in particle physics. 
Because neutrinos are electrically neutral, they can possess either Dirac mass or Majorana mass. 
The question is then "Are neutrinos Majorana or Dirac particles?", in which
the former (latter) has (no) Majorana masses with (without) lepton number violation.  
It is quite important to determine such a neutrino nature, because we can significantly narrow down possible new physics scenarios beyond the Standard Model (SM). 

One of the simplest ways to introduce the Dirac and/or Majorana masses is to add Right-Handed Neutrinos (RHNs) into the SM, 
where the latter for active neutrinos is generated via the seesaw mechanism~\cite{Minkowski:1977sc,Yanagida:1979as,Gell-Mann:1979vob,Mohapatra:1979ia,Schechter:1980gr}.
For the Dirac neutrino case, Majorana masses are forbidden by imposing the lepton number symmetry, in which a low energy effective field theory is different from the SM, because of the existence of right-handed components of the active neutrino. Currently, the Dirac neutrino case has not been excluded by experiments, so that this possibility should also be taken into account as well as the Majorana case. Although RHNs can be introduced just ``by hand", it has been known that an introduction of a new $U(1)$ gauge symmetry makes it mandatory due to the gauge anomaly cancellation. 
We thus focus on a scenario with a $U(1)_X$ symmetry~\cite{Appelquist:2002mw,Das:2017flq}, a general version of such a class of new physics including the famous $U(1)_{B-L}$ scenario~\cite{Pati:1973uk,Davidson:1978pm,Marshak:1979fm}, to naturally accommodate the RHNs. So far, models with $U(1)_X$ have been discussed in various physics motivations, e.g., collider phenomenology, flavor physics and cosmology, see~\cite{KA:2023dyz} for recent studies. 

In this Letter, we investigate the Higgs physics of the $U(1)_X$ model, and show that the discovered Higgs boson with a mass of 125 GeV can play a crucial role in new physics scenario to determine the neutrino nature. 

\noindent
{\textbf{Models}.--}  
%
\begin{table*}[t]
\begin{center}
{\renewcommand\arraystretch{1.3}
\begin{tabular}{|c||cc|cccccc|}\hline
    & $\tilde{X}_{H}$ & $\tilde{X}_{\nu_R}$& $\tilde{X}_{\Phi}$  & $\tilde{X}_{u_R}$ & $\tilde{X}_{d_R}$& $\tilde{X}_{Q_L}$ & $\tilde{X}_{L_L}$ & $\tilde{X}_{\ell_R}$   \\\hline\hline
$U(1)_X$         &$x_H^{}$         & $x_R$              & $-2x_R$ (Free)             & $\frac{1}{3}(4x_H - x_R)$ & $-\frac{1}{3}(2x_H + x_R)$ & $\frac{1}{3}(x_H - x_R)$ & $x_R - x_H$ & $x_R-2x_H$  \\\hline
\end{tabular}}
\caption{$U(1)_X$ charges $\tilde{X}_\Psi$ for 
a field $\Psi$. In the Majorana neutrino case, 
the charge for $\Phi$ is determined to be $-2x_R$ 
while it is the free parameter in the Dirac case.  
In this table, $u_R^{}$, $d_R^{}$, $e_R^{}$ are respectively the right-handed up-type quarks, down-type quarks and charged leptons, while $Q_L$ $(L_L)$ are the left-handed quarks (leptons). }
\label{tab:charges}
\end{center}
\end{table*}
We discuss a minimal extension of the SM by introducing a $U(1)_X$ gauge symmetry whose charges for SM fermions are assumed to be flavor universal and different from neither all zero (dark photon case) nor the same as the hypercharge. 
In order to satisfy the gauge anomaly cancellation, 
three RHNs $\nu_R^{i}$ ($i = 1, 2, 3$) are inevitably introduced by which tiny neutrino masses are generated. 
We take the minimal Higgs sector to realize the spontaneous symmetry breaking: 
$SU(2)_L \times U(1)_Y \times U(1)_X \to U(1)_{\rm em}$, which is composed of an isospin doublet Higgs field $H$ with the hypercharge $Y = 1/2$ and a singlet Higgs field $\Phi$ with $Y = 0$. 
Therefore, new fields introduced in our scenario are $\nu_R^i$, $\Phi$ and $Z^\prime$ with the latter being a new massive neutral gauge boson associated with $U(1)_X$ which acquires mass after the $U(1)_X$ symmetry breaking.
The $U(1)_X$ charges for all the fermions and $H$ are determined 
in terms of two free parameters by imposing the anomaly free conditions and compatibility of the Yukawa interactions for $H$
to generate usual Dirac mass terms for fermions.  
We then have two options, i.e., (i) Majorana neutrino case and (ii) Dirac neutrino case. 
For the Majorana case, additional Yukawa interactions 
\begin{align}
{\cal L}_Y \supset y_M^{ij} \overline{\nu_R^{ic}}\nu_R^j\Phi + \text{h.c.},
\label{eq:yukawa}
\end{align}
are required to give the Majorana mass term for $\nu_R^i$, and it constrains the $U(1)_X$ charge for $\Phi$ to be $-2$ times the charge for $\nu_R^{}$. 
Such a Majorana mass for $\nu_R^{i}$ generates tiny active neutrino masses via the seesaw mechanism. 
On the other hand, 
the Dirac case does not require the Yukawa interaction given in Eq.~(\ref{eq:yukawa}), so that the charge of $\Phi$ can be taken to be free.  
In this case, masses of active neutrinos are generated via the ordinal Dirac mass term as those for charged fermions.
In Table~\ref{tab:charges}, we summarize the $U(1)_X$ charges for all the fields.

In our scenario, new interactions with $Z^\prime$ 
are introduced in addition to the SM interactions. 
The covariant derivative for a field $\Psi$ is generally expressed as 
\begin{align}
D_\mu\Psi &= [\partial_\mu -ig(T_\Psi^+W_\mu^+ + {\rm c.c.}) -ieQ_\Psi A_\mu  \notag\\
&-ig_Z^{}(T^3_\Psi - s_W^2Q_\Psi)\tilde{Z}_\mu -ig_X^{}X_\Psi \tilde{Z}_\mu' ]\Psi, 
\label{eq:covariant}
\end{align}
where $T_\Psi^\pm$, $T_\Psi^3$, $Q_\Psi$ and $X_\Psi$ are respectively
the $SU(2)_L$ ladder operators, the third component of the isospin, the electric charge
and the {\it effective} $U(1)_X$ charge for $\Psi$, while $g$ and $g_X^{}$ denote the gauge couplings for $SU(2)_L$ and $U(1)_X$, respectively, and $e = g\sin\theta_W^{}$,  $g_Z^{}=g/\cos\theta_W$ with $\theta_W^{}$ being the weak mixing angle. We note that effects of the kinetic-mixing between $U(1)_Y$ and $U(1)_X$
are already taken into account in Eq.~(\ref{eq:covariant}), by which 
the relation between the effective charge $X_\Psi$ and the original charge $\tilde{X}_\Psi$ is given by 
$X_\Psi = \tilde{X}_\Psi - Y_\Psi(g/g_X^{})\tan\theta_W^{}\tan\epsilon$
with $\epsilon$ being the kinetic-mixing parameter. 
Throughout this Letter, we take $\epsilon = 0$ for simplicity.  
In Eq.~(\ref{eq:covariant}), $W_\mu^\pm$ ($A_\mu$) are the W bosons (photon). Two neutral massive bosons $\tilde{Z}_\mu$ and $\tilde{Z}_\mu^\prime$
are generally mixed with each other via the angle $\zeta$
\begin{align}
\begin{pmatrix}
\tilde{Z}_\mu\\
\tilde{Z}_\mu^{\prime}
\end{pmatrix}
=
R(\zeta)
\begin{pmatrix}
Z_\mu\\
Z^\prime_\mu
\end{pmatrix},~~
R(\theta) = \begin{pmatrix}
\cos\theta & -\sin\theta \\
\sin\theta & \cos\theta
\end{pmatrix}, 
\end{align}
where 
$\sin2\zeta = g_Z^{}v^2\delta/(m_{Z'}^2 - m_Z^2)$
with $\delta \equiv g_X^{}X_H$.
As we will see below, the magnitude of $\sin\zeta$ or $\delta$ has to be much smaller than unity due to constraints from current experimental data such as the electroweak rho parameter $\rho$. Therefore, it is convenient to deal with $\delta$ as a perturbation parameter. 
Under $\delta \ll 1$, the masses of $Z$ and $Z'$ are expressed as 
\begin{align}
m_Z^2  &= \frac{g_Z^2}{4}v^2 + \frac{g_Z^2 v^4 \delta^2}{g_Z^2v^2 - 4g_X^2X_\Phi^2 v_\Phi^2} + {\cal O}(\delta^4) , \label{mZp}\\
m_{Z'}^2 &= g_X^2X_\Phi^2v_\Phi^2   -\frac{4g_X^2 X_\Phi^2 v^2v_\Phi^2 \delta^2}{g_Z^2v^2 - 4g_X^2X_\Phi^2 v_\Phi^2} + {\cal O}(\delta^4).
\label{eq:mzpsq}
\end{align}

The most general form of the Higgs potential is expressed as 
\begin{align}
V&= -\mu_H^2 |H|^2 -\mu_\Phi^2|\Phi|^2 
+ \frac{\lambda_H}{2}|H|^4 + \frac{\lambda_\Phi}{2}|\Phi|^4\notag\\ 
&+ \lambda_{H\Phi}|H|^2|\Phi|^2. 
\end{align}
The $\lambda_{H\Phi}$ term induces a mixing between physical neutral components of $H$ ($\tilde{h}$) and $\Phi$ ($\tilde{\phi}$) as $(\tilde{h}, \tilde{\phi} )^T=R(\alpha) (h, \phi)^T $
where $h$ ($\phi$) can be identified with the discovered Higgs boson (an extra Higgs boson) and $\alpha$ is the mixing angle. 
The five parameters in the potential can then be expressed by the physical parameters $\{v,v_\Phi^{},m_h,m_\phi,\sin\alpha\}$, 
where $m_h$ and $m_\phi$ are the masses of $h$ ($m_h \simeq 125$ GeV) and $\phi$ respectively, and 
$v \equiv \sqrt{2}\langle H^0 \rangle \simeq 246$ GeV and $v_\Phi \equiv \sqrt{2}\langle \Phi^0 \rangle$.

\noindent
{\textbf{Decays of $Z^\prime$ and $h$.--} }
%
For the later discussion, we consider the decay properties of $Z^\prime$ and $h$. 
We particularly concentrate on the 
case for $m_\phi > m_h,~m_{Z'}$, in which 
$h$ and $Z^\prime$ do not decay into a pair of $\phi$. 

\begin{figure}
\includegraphics[scale=0.36]{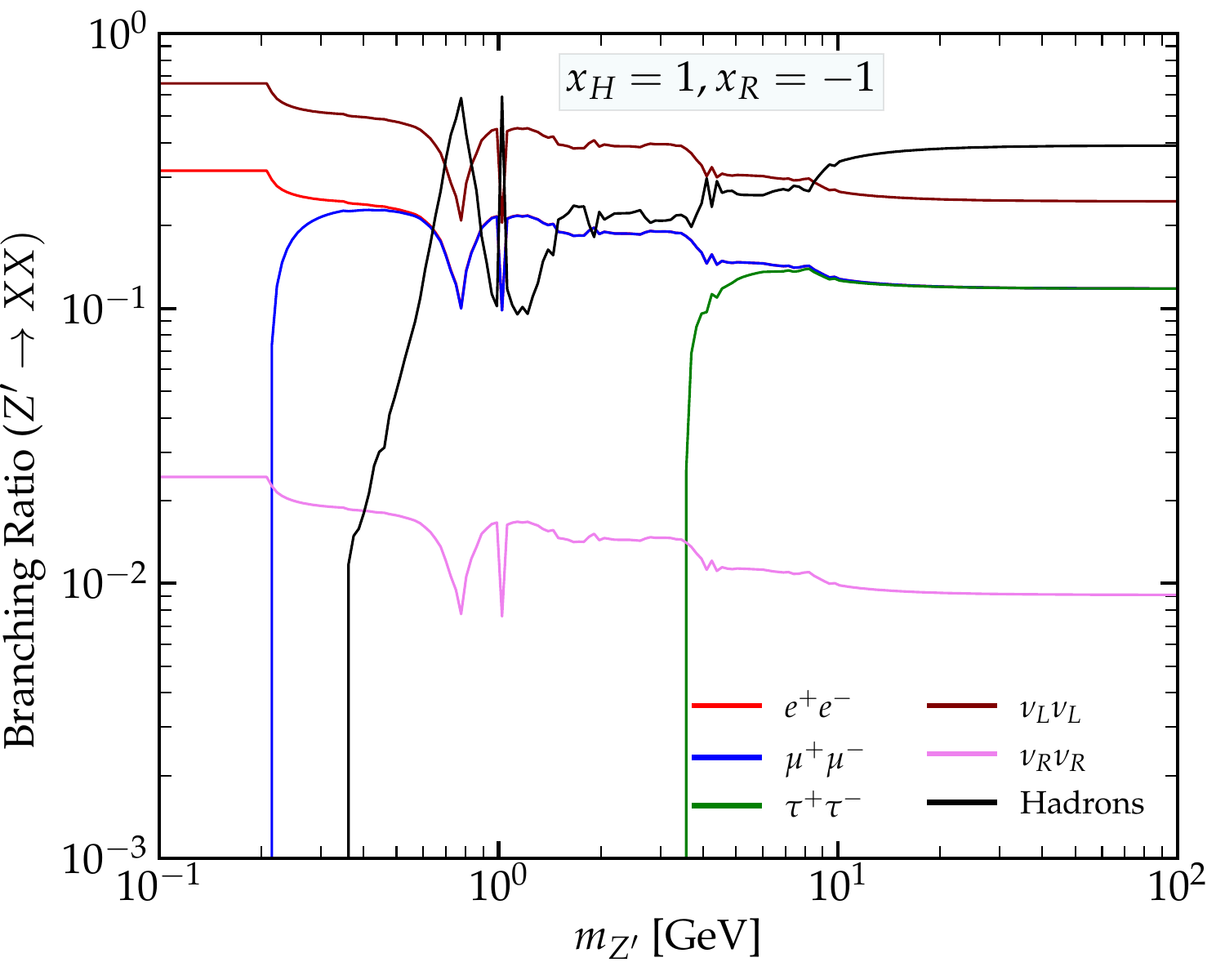}
\caption{Branching ratio of $Z^\prime$ as a function of $m_{Z^\prime}$ for $x_H=1$ and $x_R=-1$.}
\label{fig:br_zp}
\end{figure}

The decay of $Z^\prime$ into a fermion pair $Z^\prime \to f\bar{f}$ depends on the $U(1)_X$ charges for 
a left-handed fermion ($X_{f_L}$) and a right-handed fermion ($X_{f_R}$). Its decay rate at parton level is given by 
\bea
\label{eq:zpffbar}
\Gamma(Z^\prime \to f \bar{f}) &=& \frac{m_{Z^\prime} g_X^2 N_c^f}{24 \pi} \left[\left( X_{f_L}^2 + X_{f_R}^2 \right) \left(1 - \frac{{m_{f}^2}}{{m_{Z^\prime}^2}}\right) \right. \nonumber \\
 &&\left. + 6 X_{f_L} X_{f_R} \frac{m_f^2}{m_{Z^\prime}^2} \right] \sqrt{1 - \frac{{4 m_{f}^2}}{{m_{Z^\prime}^2}}}, 
\eea
where $m_f$ is the mass of the fermion and $N_c=1~(3)$ is the color factor for leptons (quarks).
In Fig.~\ref{fig:br_zp}, we show the branching ratios of $Z^\prime$
as a function of $m_{Z^\prime}$, which 
is evaluated by using the {\tt darkcast} package~\cite{Ilten:2018crw} including effects of the hadronization considering $x_H=1$ and $x_R=-1$ where all the fermions couple with $Z^\prime$.
We take the Majorana mass for RNHs to be the typical 
active neutrino mass, i.e., 0.1 eV, so that there is no difference between the Dirac and Majorana cases in the decay of $Z^\prime$ into neutrinos.   
We see that the branching ratio of 
the ``golden channel" $Z^\prime \to e^+e^-$ or $Z^\prime \to \mu^+\mu^-$ is about 10\%.  

The Higgs boson $h$ can decay into new final states, i.e., 
$h \to Z'Z'$, $h \to Z'Z$, $h \to Z'\gamma$ and $h \to \phi\phi$, where the last one is kinematically forbidden in the current setup. The decay rates for the first two modes are given up to the term ${\cal O}(\delta^2)$ by 
\begin{align}
&\Gamma(h \to Z'Z') = \frac{m_h^3}{32\pi v_\Phi^2}\left[s_{\alpha}^2 + \frac{2m_{Z'}^2v^2\delta^2}{(m_Z^2 - m_{Z'}^2)^2}\right. \notag\\
&~~\left.\left(s_\alpha^2 - \frac{v_\Phi^{}}{v}c_\alpha s_\alpha \right) \right]\left(1 - 4x_{Z'}^{} + 12x_{Z'}^2\right)(1 - 4x_{Z'}^{}), \label{eq:hzpzp} \\
&\Gamma(h \to ZZ') = \frac{m_h^3}{16\pi}\frac{m_{Z'}^2\delta^2}{(m_Z^2 - m_{Z'}^2)^2}\left(\frac{vs_\alpha}{v_\Phi} - c_\alpha \right)^2 \notag\\
&~~\left[\lambda(x_Z^{},x_{Z'}^{}) + 12x_Z^{}x_{Z'}^{}\right]\lambda^{1/2}(x_Z^{},x_{Z'}^{}),
\end{align}
where $x_i = m_i^2/m_h^2$ and $\lambda(x,y)=(1-x-y)^2-4xy$. We see that the decay rate of $h\to ZZ'$ is suppressed by $\delta^2$, so that it is typically much smaller than that of $h \to Z'Z'$ as long as $s_\alpha^2 \gg \delta^2$. 
The decay of $h \to Z'\gamma$ is further suppressed by the loop-factor squared with respect to $h \to ZZ'$, so that we can safely neglect it. 
We note that the first and second terms in the square bracket in Eq.~(\ref{eq:hzpzp}) are respectively given by the effect of the Higgs mixing and the interference between
the Higgs mixing and the $Z$-$Z^\prime$ mixing where the latter is proportional to $\delta^2$. 
Therefore, as far as we neglect the ${\cal O}(\delta^4)$ term, the decay rate of $h \to Z^\prime Z^\prime$ vanishes at $s_\alpha = 0$. 
%
For $\delta \ll 1$ and $m_{Z'}/m_h \ll 1$, Eq.~(\ref{eq:hzpzp}) becomes
\begin{align}
\Gamma(h \to Z'Z') \sim \frac{m_h^3s_{\alpha}^2}{32\pi v_\Phi^2},  \label{eq:hzpzp2}
\end{align}
which can also be derived by using the equivalence theorem, i.e., $\Gamma(h \to Z'Z') \simeq \Gamma(h \to G_{Z'}G_{Z'})$ with $G_{Z'}$ being the Nambu-Goldstone boson associated with $Z'$.  
It is clear that the decay rate of the $h \to Z'Z'$ process is simply determined by two free parameters $\sin\alpha$ and $v_\Phi^{}$ and is enhanced for smaller $v_\Phi^{}$.
Since the $Z'$ mass is roughly determined by $g_X^{}X_\Phi v_\Phi$, see Eq.~(\ref{eq:mzpsq}), smaller $v_\Phi$ can be obtained by taking larger $X_\Phi$ under a fixed value of $m_{Z'}/g_X^{}$.
Therefore, a sizable value of the branching ratio of $h \to Z'Z'$ can be obtained in the {\it Dirac neutrino} case as compared with the Majorana neutrino case, because $X_\Phi$ can be taken to be free in the former, see Table~\ref{tab:charges}.

\begin{figure}
\centering
\includegraphics[scale=0.35]{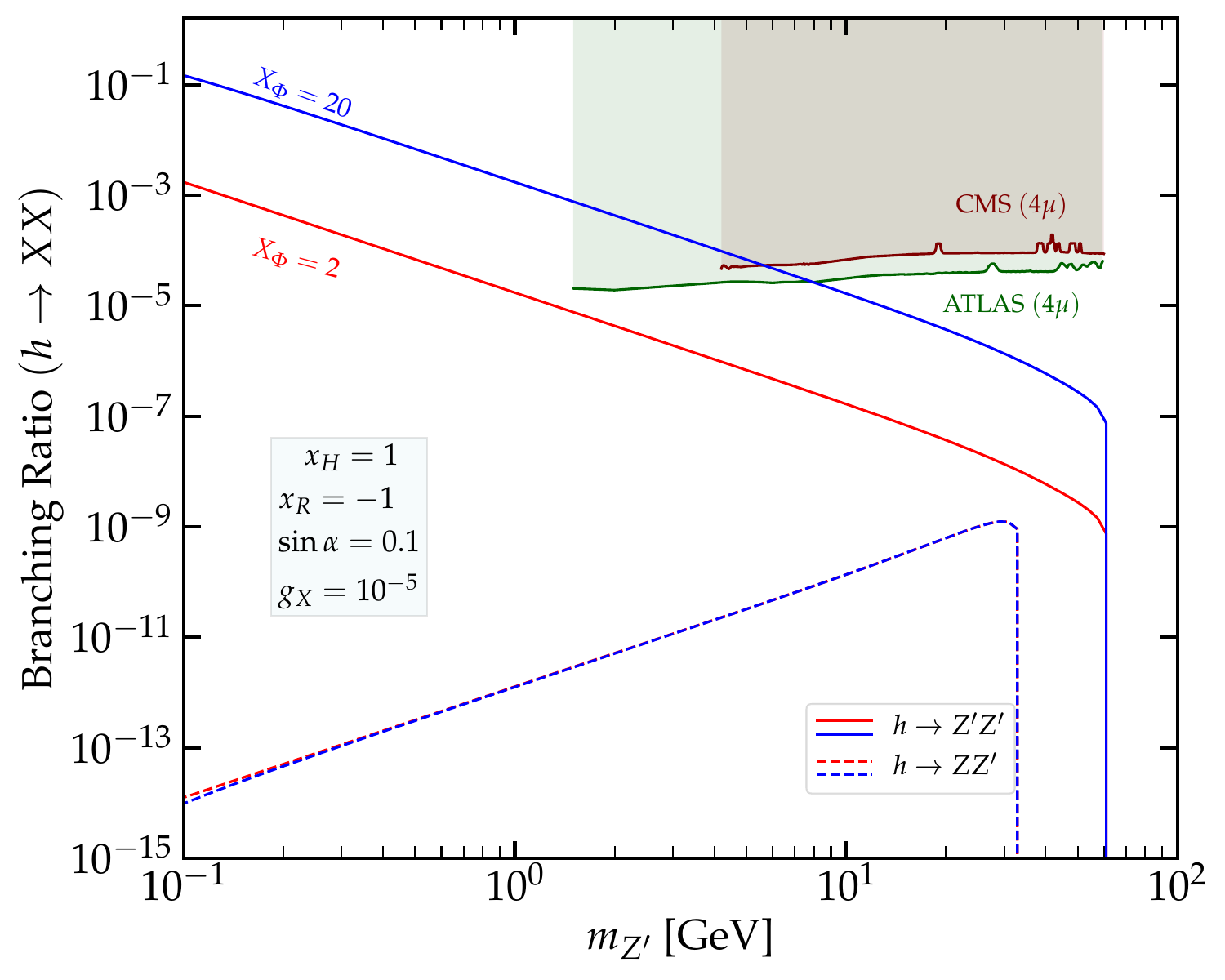}
\caption{Branching ratios of $h \to Z^\prime Z^\prime$ (solid) and $h\to ZZ^\prime$ (dashed) as a function of $m_{Z^\prime}$ with  $X_{\Phi}=2$ (Majorana and Dirac cases) and 20 (Dirac case only).  The shaded region is excluded by the $4\mu$ search at LHC, see the text for details. }
\label{fig:br_higgs}
\end{figure}

In Fig.~\ref{fig:br_higgs}, we show the branching ratios of $h$ into the $Z Z^\prime$ and $Z^\prime Z^\prime$ modes as a function of $m_{Z^\prime}$. 
We here take $g_X^{} = 10^{-5}$ as a typical value allowed by the current experimental data.
These branching ratios drop sharply at the vicinity of $m_h/2$ for $h \to Z^\prime Z^\prime$ and $m_h - m_Z^{}$ for $h \to ZZ^\prime$ due to the kinematical threshold.  
As we expect, BR($h \to ZZ'$) is much smaller than BR($h \to Z'Z'$) due to the large suppression by $\delta^2$. We also see that $h \to Z'Z'$ mode is enhanced for smaller $m_{Z^\prime}$ corresponding to the smaller value of $v_\Phi^{}$. 
In addition, a larger value of BR($h \to Z'Z'$) is obtained for a larger value of $X_\Phi$, which is possible in the Dirac case, while $X_\Phi$ is fixed to be 2 in the Majorana case. 

\noindent
{\textbf{Constraints.--}} 
We take into account the following constraints on the parameter space, e.g., $\{g_X^{},~m_{Z^\prime},~s_\alpha\}$,  given by the existing data.

\noindent
{\it (a) Higgs signal strength at LHC:} 
The latest value 
for the inclusive signal strength $\mu^{}$ 
is given by $\mu = 1.05\pm 0.06$~\cite{ATLAS:2022vkf}. 
In our scenario, we can simply estimate the $\mu$ value to be $c_\alpha^2$, because the branching ratios of $h$ into SM particles are almost the same as the SM prediction, and the production cross section for $h$ is modified by the factor of $c_\alpha^2$. 
Thus, we obtain the upper limit on the Higgs mixing angle as $|s_\alpha| \lesssim 0.264$ at 2$\sigma$ level.


\noindent
{\it (b) $\rho$-parameter:} The $Z$-$Z^\prime$ mixing causes the electroweak rho parameter  different from unity at tree level. For $\delta \ll 1$, $\rho$ is given at tree level by 
\begin{align}
\rho\simeq 1 - \frac{v^2\delta^2}{m_Z^2 - m_{Z'}^2},
\end{align}
so that $|\delta| \lesssim 9.79\times 10^{-3}$ (for $m_{Z'}/m_Z \ll 1$) is obtained at 95\% CL from the measurement $\rho_{\rm exp} = 1.0003\pm 0.0005$~\cite{Workman:2022ynf}. 

\noindent
{\it (c) Dark photon searches at flavor experiments:} We estimate bounds on $\{m_{Z'}, g_X \}$ from the dark photon $(A^\prime)$ searches with the decay $A^\prime \to \mu^+ \mu^-$ at the LHCb~\cite{LHCb:2019vmc} and CMS~\cite{CMS:2023slr} experiments. 
These experiments have put the limit on the dark photon mass and the kinetic mixing parameter, which can be converted into 
the constraint on $g_X^{}$ and $m_{Z^\prime}$. 
We apply the conversion method given in Ref.~\cite{KA:2023dyz} to our analysis. 
For a fixed value of $m_{Z^\prime}$, we obtain the lower (upper) limit on $\tilde{v}_\Phi^{} \equiv v_\Phi^{} X_\Phi$ ($g_X^{}$). For instance, we obtain $\tilde{v}_\Phi > 46.9$ TeV corresponding to $g_X^{} < 3.2\times 10^{-5}$ for $m_{Z^\prime}=1.5$ GeV from LHCb. 

\noindent
{\it (d) Cosmological constraints:} In general, new relativistic particles can contribute to the effective number of neutrinos $N_{\rm eff}$, because they can change the energy density of radiations in the early Universe. In our scenario with the Dirac neutrino case, the mass of $\nu_R^{}$ is taken of the order of active neutrino mass. In addition to that, $\nu_R^{}$ can have a sizable interaction with $Z^\prime$ lowering its decoupling temperature up to that of active neutrinos, i.e., $\mathcal{O}$(1) MeV. In such a case, we have to take into account the constraint from $N_{\rm eff}$. 
According to Ref.~\cite{Luo:2020sho}, an effective coupling for four neutrino scatterings $G_V$ defined via the effective Lagrangian 
\begin{align}
{\cal L}_{\rm eff}= G_V (\bar{\nu}_L^{} \gamma^\mu \nu_L^{})(\bar{\nu}_R^{} \gamma_\mu \nu_R^{}), \label{eq:eff}
\end{align}
is constrained to be $|G_V| < (11.4~\text{TeV})^{-2}$ at 95\%~\text{CL}
by imposing the current measurement of $N_{\rm eff}$ 
at Planck 2018, i.e., $N_{\rm eff} = 2.99\pm 0.17$~\cite{Planck:2018nkj}.
In our scenario, the $Z'$ mediation gives rise to the effective interaction (\ref{eq:eff}), and we can identify 
\begin{align}
G_V = -\frac{X_{L_L}X_{\nu_R}}{\tilde{v}_\Phi^2}. \label{eq:gv2}
\end{align}
We thus obtain the lower limit on $\tilde{v}_\Phi^{}$ to be of order 10 TeV for 
$X_{L_L}X_{\nu_R} \sim {\cal O}(1)$.

\noindent
{\textbf{Results.--}} 
Let us discuss the $Z^\prime$ search 
via the Higgs decay $h \to Z^\prime Z^\prime$ at LHC and future collider experiments.  
In particular, we consider the $h \to Z^\prime Z^\prime \to 4\mu$ process, by which we can exclude and/or explore the largest region of the parameter space in our scenario.
The cross section for the four muon process is calculated as 
\bea
\sigma_{4\mu} = \sigma_{h}^{\rm SM}\times c_\alpha^2\times {\rm BR}(h \to Z^\prime Z^\prime)\times  {\rm BR}(Z^\prime \to \mu\mu)^2,~~~~
\label{limits}
\eea
where $\sigma_{h}^{\rm SM}\simeq 55.73$ pb is the production cross section for the SM Higgs boson at LHC with the collision energy of 13 TeV~\cite{ATLAS:2020rej}. 
The ATLAS and CMS collaborations have independently performed
the $Z^\prime$ search via the Higgs decay by using Run-II data, and they have put the model independent upper limit on the cross section $\sigma_{4\mu}$ to be about 0.03 fb~\cite{ATLAS:2021ldb} and 0.05 fb~\cite{CMS:2021pcy} at 95\% CL, respectively, for $m_{Z^\prime}=5$ GeV. Similar values of the upper limit have been taken for the other choices of $m_{Z^\prime}$ up to $m_h/2$. 
In the following analysis, we take into account the ATLAS limit only 
as the CMS limit is slightly weaker than the ATLAS one.
We can convert the upper limit on $\sigma_{4\mu}$ into the upper limit on BR($h \to Z'Z'$) by using our prediction of BR($Z^\prime \to \mu\mu$). 
For instance, we find that BR$(h\to Z^\prime Z^\prime)$ has to be smaller than  
$\mathcal{O}(10^{-5})$ for $x_H=1$, $x_R = -1$, $s_\alpha=0.1$ and $g_X=10^{-5}$ in the range of $1.5$ GeV $\leq m_{Z^\prime} \leq m_h/2$. 
We can also extract the future sensitivity for the $Z^\prime$ search at High-Luminosity LHC (HL-LHC) by extrapolating the ATLAS limit as
\begin{align}
\sigma_{4\mu}^{\rm HL-LHC} = \sigma_{4\mu}^{\rm Run-II}\sqrt{139/3000}, 
\end{align}
with $\sigma_{4\mu}^{\rm Run-II}$ being the upper limit on $\sigma_{4\mu}$ given by the Run-II data with the integrated luminosity of 139 fb$^{-1}$~\cite{ATLAS:2021ldb}. 

In addition to the expectation at HL-LHC, we consider precise measurements 
for the Higgs boson couplings at International Linear Collider (ILC), especially the one with the collision energy of 250 GeV and the integrated luminosity of 2 ab$^{-1}$ (ILC250).  The $hZZ$ coupling is expected to be measured with an accuracy of 0.38\% at 1$\sigma$ level based on the so-called $\kappa$-framework~\cite{Fujii:2017vwa}, and it can be translated into a limit $|s_\alpha| \lesssim 0.123$ assuming the central value of the $hZZ$ coupling is measured to be the same value as the SM prediction. 

\begin{figure}
\centering
\includegraphics[scale=0.47]{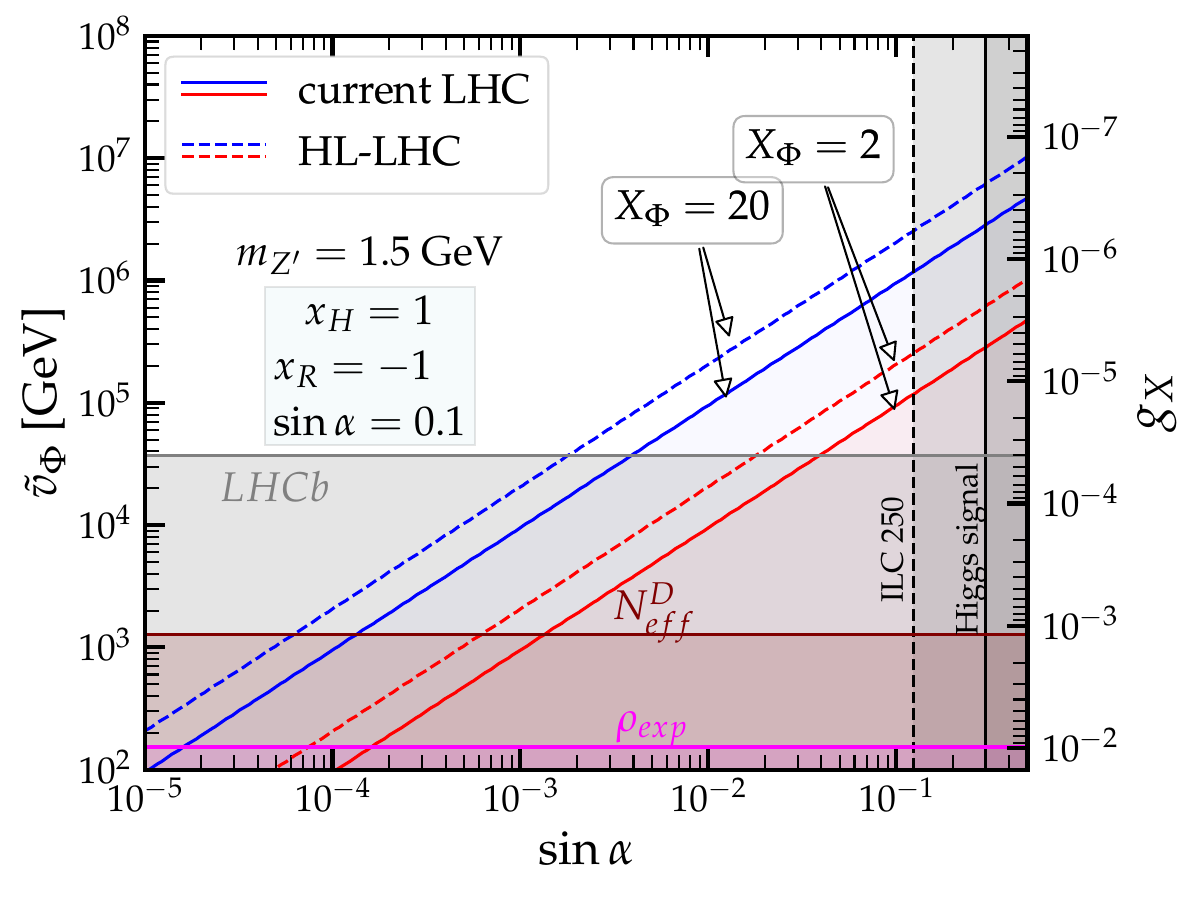}
\caption{Region excluded by the current experimental data and that 
excluded and/or explored by the future experiments on the $s_\alpha$-$\tilde{v}_\Phi$ plane.
We fix $m_{Z^\prime}=1.5$ GeV, $x_H=1$ and $x_R = -1$, which determines $X_\Phi^{}$ to be 2 in the Majorana case. For the Dirac case, we take $X_\Phi=2$ and 20. 
The shaded region is excluded by the current experimental data, while the dashed lines 
show the future sensitivities.  
} 
\label{fig:br_higgs-1}
\end{figure}


Now, let us combine all the current constraints and the future sensitivities
discussed above. 
In Fig.~\ref{fig:br_higgs-1}, we show the parameter space excluded by the constraints on the $s_\alpha$-$\tilde{v}_\Phi^{}$ plane. 
We also show the corresponding value of $g_X^{}$ on the right side of the vertical axis. 
The left region from the vertical black line is excluded by the Higgs signal strength (a), while that below the gray, brown and magenta solid lines is excluded by the constraints from the flavor data (c), cosmological constraint (d) and the rho parameter (b), respectively. 
We note that the region excluded by the above constraints 
is not changed if we take a different value of $X_\Phi$ (in any case, it is fixed to be 2 in the Majorana case). 
On the other hand, the constraint from the four muon search at LHC
strongly depends on the choice of $X_\Phi$. 
The region below the red and blue solid line is respectively excluded for 
$X_\Phi = 2$ and $X_\Phi = 20$, and the latter choice is only possible in the Dirac case. 
We also show the region which can be excluded or explored at HL-LHC (red and blue dashed line) and at ILC (black dashed line). 
It is clear that we can exclude or explore only the corner of the parameter space in the Majorana case, but the large portion of the parameter space can be explored in the Dirac case, i.e., varying the value of $X_\Phi$ from 20, 
the blue solid and dashed lines move parallel to the orthogonal direction of each line.  

Finally, we show the constraint on the parameter space on the $m_{Z^\prime}$-$g_X^{}$ plane. 
The gray region is excluded by the flavor data (c), while 
the region above the red and blue solid (dashed) curves are excluded (excluded or explored) by the four muon search at the current LHC (at HL-LHC) with $X_\Phi = 2$ and $X_\Phi = 20$ (Dirac case only), respectively. 
As compared with the Majorana case, 
we can exclude or explore the case with 
much smaller values of $g_X^{}$ 
than the current upper limit given by the flavor data
in the Dirac case.


\begin{figure}
\centering
\includegraphics[scale=0.33]{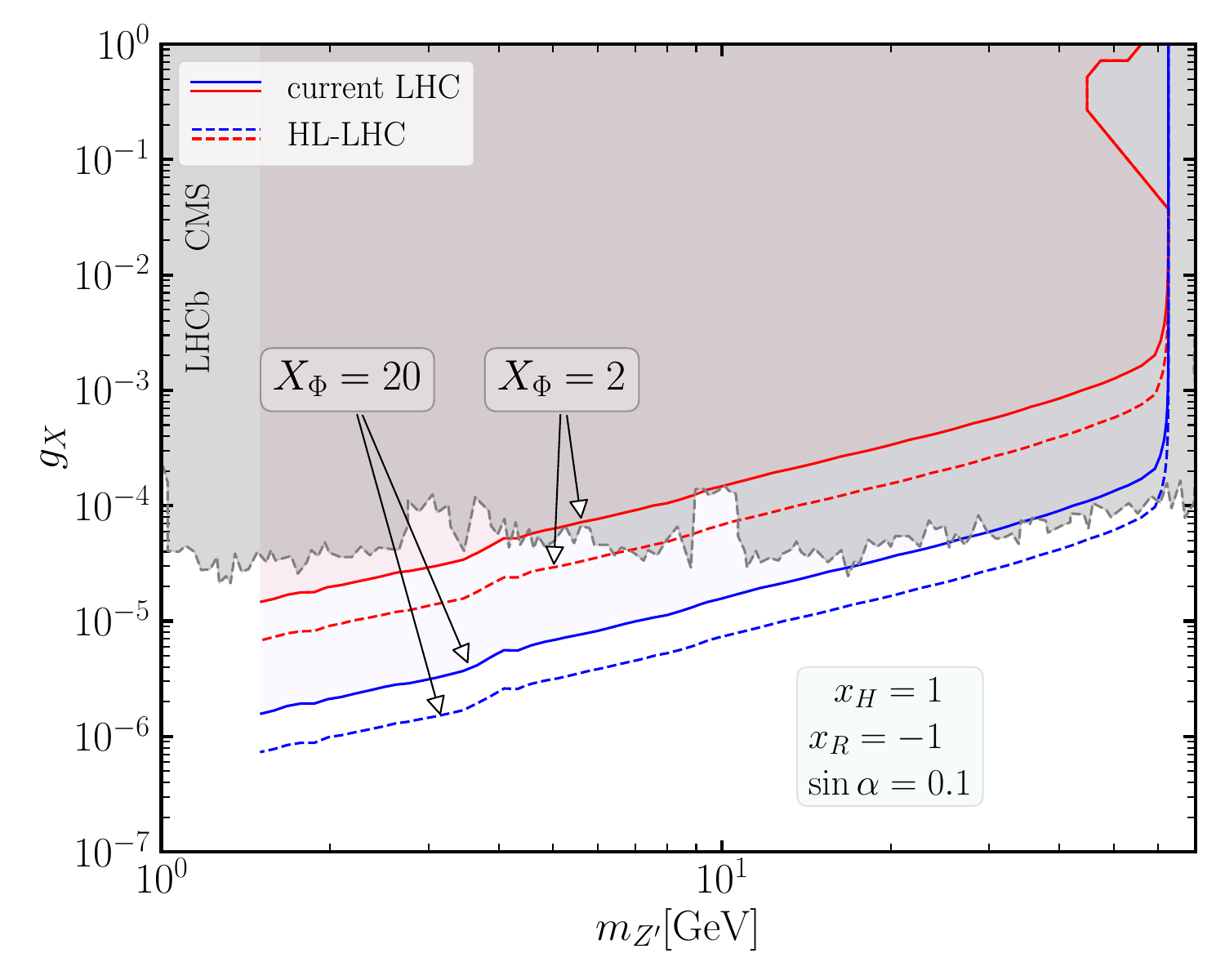}
\caption{Similar to Fig.~\ref{fig:br_higgs-1}, but the constraints are shown on the $m_{Z^\prime}$-$g_X$ plane. }
\label{fig:mz_gx}
\end{figure}

\noindent
{\textbf{Conclusions.--}} We have considered two different $U(1)_X$ scenarios where active neutrinos are regarded as Majorana and Dirac particles. 
The crucial difference between these two scenarios appears in the $U(1)_X$ charges, i.e., the charge for the singlet Higgs field $\Phi$ can be taken to be free in the Dirac case, while it is determined by the other charges in the Majorana case.   
Because of this additional degree of freedom, 
the $h \to Z^\prime Z^\prime$ decay can be significantly enhanced in the Dirac case as compared with the Majorana case. 

We have taken into account the constraint from the searches for the $h \to Z^\prime Z^\prime \to 4\mu$ channel at current LHC in addition to the known bounds, e.g., flavor constraints (LHCb), the electroweak precision data (rho parameter), the Higgs signal strength and the cosmological constraint (the effective number of neutrinos $N_{\rm eff}$).   
We have found that the constraint from the $4\mu$ channel
can exclude a portion of the parameter space which is not excluded by the known previous constraints. 
In the Majorana case, such a newly excluded region is quite limited due to the strong constraint by the LHCb data.  
On the other hand, in the Dirac case, a much wider region of the parameter space can be excluded by the $4\mu$ search due to the enhancement of the $h \to Z^\prime Z^\prime$ decay.  
Therefore, the Higgs decay can be a powerful tool to prove the model with the $U(1)_X$ gauge symmetry, especially in the case where active neutrinos possess the Dirac-type mass, by which we can indirectly probe the neutrino nature in future.


\noindent
{\textbf{Acknowledgments.--}} SKA thanks the Department of Ministry of Education, Culture, Sports, Science and Technology of Japan for MEXT fellowship to study in Japan. This work was supported in part by JSPS KAKENHI Grants Nos. 20H00160, 22F21324 and 23K17691.
\bibliographystyle{utphys}
\bibliography{references}
\end{document}